\documentclass[rnote]{aa} 
\usepackage{graphicx}
\usepackage{natbib}
\usepackage{txfonts}
\usepackage[dvips,usenames]{color}
\bibpunct{(}{)}{;}{a}{}{,} 

\newcommand{\realfigure}[3]{
              \begin{figure}
              \includegraphics[width=8.5cm]{#1}
              \caption{#2}\label{#3}
              \end{figure}
              }

\begin{document}
   \title{Stellar models with the ML2 theory of convection
	}


   \author{M. Salaris \inst{1}     \and S. Cassisi \inst{2} }

   \offprints{M. Salaris}

   \institute{
              Astrophysics Research Institute, Liverpool John Moores
              University, Twelve Quays House, Egerton Wharf Birkenhead
              CH41, 1LD, UK \\ 
              \email{ms@astro.livjm.ac.uk}
         \and 
              INAF - Osservatorio Astronomico di Collurania, 
              Via Mentore Maggini, I-64100 Teramo\\
              \email{cassisi@oa-teramo.inaf.it}
          }

\date{}


\abstract
{The mixing length theory (MLT) used to compute 
the temperature gradient in superadiabatic layers of stellar (interior and atmosphere) models 
contains in its standard form 4 free parameters. Three parameters are fixed a priori 
(and define what we denote as the MLT 'flavour') whereas one (the so-called mixing length) is 
calibrated by reproducing observational constraints. The \lq{classical}\rq  B\"ohm-Vitense flavour 
is used in all modern  
MLT-based stellar model computations and, despite its crude approximations, 
the resulting $T_{eff}$ scale appears -- perhaps surprisingly -- remarkably realistic, once 
the mixing length parameter is calibrated with a solar model.} 
{Model atmosphere computations employ parameter choices different from what is used in 
stellar interior modelling, raising the question of whether a single MLT 
flavour and mixing length value can be used to compute interiors and atmospheres of stars of all types. 
As a first step towards addressing this issue, we study whether the MLT flavour (the so-called ML2) 
and mixing length choice that have been proven adequate to model white dwarf atmospheres, is able to provide, 
when used in stellar models, results at 
least comparable to the use of the \lq{classical}\rq  B\"ohm-Vitense flavour.}
{We have computed solar models and evolutionary tracks for both low- and intermediate-mass Population I and II  
stars, adopting both solar calibrated B\"ohm-Vitense and ML2 flavours of the MLT in our stellar evolution code, 
and state-of-the-art input physics. }
{The two sets of models 
provide consistent results, with only minor differences. Both calibrations reproduce also the $T_{eff}$ of 
red giants in a sample of Galactic globular clusters. The ML2 solar model  
provides a mixing length about half the value of the local pressure scale height, thus alleviating -- but 
not eliminating -- one of the well known inconsistencies of the MLT employed in stellar models. This mixing length is also 
consistent with the value used in white dwarf model atmosphere 
computations.}{}


   \maketitle
\section{Introduction}

The equations still largely employed 
to determine the value of the temperature gradient in 
the superadiabatic regions of stellar convective envelopes and atmospheres  
are based on the so called ``mixing length theory'' (MLT). 
The MLT in its original form is a simple, local, time independent model, firstly 
applied to stellar modelling by \citet{bie32}.  
Its description of convective transport considers the motion of 'average'
convective cells, all with the same physical properties at a given radial
location $r$ within the star. These convective cells have a mean size $l$ 
($l=\alpha_{MLT} H_P$ is the so-called 'mixing length', 
where $\alpha_{MLT}$ is a free parameter and $H_P$ is 
the local pressure scale height $H_P\equiv P/g \rho$) also equal to their
mean free path, and an average convective speed
${\rm v_c}$ at a given value of $r$. 
There is in principle no reason why $\alpha_{MLT}$ should be kept constant when considering 
stars of different
masses and/or chemical composition and/or at different evolutionary stages; even within
the same star $\alpha_{MLT}$ might in principle vary from layer to layer, although
stellar models usually keep $\alpha_{MLT}$ constant throughout the convective zone 
\citep[but see][for some low-mass stellar 
models computed with varying $\alpha_{MLT}$ within the convective envelope]{ped90}.
Keeping $\alpha_{MLT}$ constant still means a change of the cells' mean size and free path due to
the variation of $H_p$.

The value of $\alpha_{MLT}$ is usually calibrated by reproducing some empirical
constraint. In case of stellar models, it is typically the calibration 
of a theoretical solar model 
that determines the value of $\alpha_{MLT}$. In all major stellar model databases  
\citep[e.g.][]{g00, vdb00, yi01, pcsc04, d07} 
this solar calibrated $\alpha_{MLT}$ is kept fixed in all evolutionary 
calculations of stars of different masses and chemical compositions.
There are 3 additional free parameters entering the MLT  
that are generally fixed a priori, before the calibration of $\alpha_{MLT}$. 
They will be denoted here as $a, b, c$ following the formalism 
of \citet{tas90}. These three parameters plus $\alpha_{MLT}$ enter the equations that determine 
${\rm v_c}$, the average convective flux 
$F_c$ and the convective efficiency $\Gamma$ (defined as 
the ratio between the excess heat content of a raising convective bubble just before its 
dissolution, and the energy radiated during its lifetime)
at each point within the convective envelope. Once $a, b, c$ are fixed, the temperature gradient at a 
given location in the convective envelope of 
a stellar model and the resulting $T_{eff}$, depend on the value of $\alpha_{MLT}$.
On the other hand, when $\alpha_{MLT}$ is kept fixed and one or more of the other three parameters 
are varied, the temperature gradients and $T_{eff}$ are also affected 
\citep[e.g.][]{heny65}.

The 'classical' formulation of the MLT by \citet{bv58} that is nowadays almost 
universally used in theoretical computations of  
stellar evolution models (denoted here as the BV58 'flavour' of the MLT) 
employs a specific set of choices for the values for $a, b$ and $c$. 
Solar calibrations with the present generation of stellar input physics 
provide a value of $\alpha_{MLT}$ typically around $\sim$2.0   
\citep[e.g.][]{vdb00, pcsc04, bsb06, d07, ml08} the precise value (for a given set of input physics) 
being dependent on the choice for the boundary conditions at the model surface. 

The situation for model atmosphere computations is different. The 
model grids perhaps most used to produce synthetic spectra and 
bolometric corrections  
for evolutionary phases before the White Dwarf (WD) stage are 
the ATLAS9, MARCS and PHOENIX ones. The recent ATLAS9 grid described in \citet{pcsc04} employs 
$\alpha_{MLT}$=1.25, while the latest MARCS models \citep{gu08} adopt a different value 
$\alpha_{MLT}$=1.5. There are some differences between the MLT flavours 
used in thse two sets of models, and also compared to the BV58 formalism used in 
stellar evolution computations \citep[see][]{cas97,gu08}.
ATLAS9 models with a much smaller $\alpha_{MLT}$=0.5 have also been computed 
\citep{hei02}.
The recent \citet{bh05} PHOENIX grid employs $\alpha_{MLT}$=2.0 
and (as communicated to us by our referee) the BV58 flavour of the MLT.

Yet another flavour of the MLT is employed in calculations of 
model atmospheres, spectra and bolometric corrections for WD stars \citep[e.g.][]{bwb95, rohr01}. 
It is the so-called ML2 \citep{tas90}, with its own specific choice of $a, b$ and $c$, 
different from the case of BV58. \citet{bwl95} have shown how for the case of ZZ~Ceti stars, 
the ML2 with $\alpha_{MLT}$=0.6 provides overall consistency between temperature estimates from both UV 
and optical spectrum, observed photometry, gravitational redshift mass estimates and 
trigonometric parallax. The exact form of the MLT employed in WD interior models 
seem to be irrelevant \citep[see, e.g.][]{fbb01} so that effectively one can employ the ML2 in computations 
of both convective atmospheres and envelopes of WDs.

This variety of choices regarding the 4 free parameters of the MLT should probably not be surprising. 
As already stated before, given the very approximate nature of the MLT, one cannot a 
priori expect the same choice of $\alpha_{MLT}$ -- and possibly also of $a, b$ and $c$ -- to be appropriate for 
all evolutionary phases in both convective envelope and atmosphere of stars. It is however interesting, 
if only for heuristic purposes, to investigate whether a single 
flavour of the MLT with just one choice (or a small range of values) 
of $\alpha_{MLT}$ is adequate 
to model convective envelopes and atmosphere 
covering all major evolutionary phases. Here we present a first step in this direction, 
where we have selected as reference choice the ML2.  
Our aim is to study whether a solar calibration with the ML2 provides 
a value of $\alpha_{MLT}$ close to what is 
employed in WD studies, how this 'new' solar model compares with a BV58 solar model, and how 
evolutionary tracks computed with the solar calibrated ML2 compare with the BV58 calibration and with 
empirical determinations of $T_{eff}$ for Red Giant Branch (RGB) stars in Galactic globular clusters.
Based on the results by \citet{gou76} and \citet{ped90} we should expect that stellar models computed 
with different choices for $a$, $b$ and $c$ are largely equivalent once $\alpha_{MLT}$ is 
appropriately recalibrated (although this has not yet been verified along the upper RGB) 
but we definitely do not know whether the same value of $\alpha_{MLT}$ 
employed in WD model atmosphere studies is adequate also for stellar modelling with the ML2. 

We wish to stress that the aim of this note is not to argue for the superiority of the MLT over 
alternative and less crude prescriptions for stellar convection \citep[see, e.g.][]{canu91}, 
rather to investigate whether, within the framework of the MLT -- 
that in spite of all its intrinsic limitations is still widely used 
in stellar evolution and model atmosphere calculations -- and with 
state-of-the-art physics inputs, it is possible to find a single combination of 
parameters suitable for computing interior and atmosphere models of all types of stars 
with superadiabatic convective layers. With 'suitable' we mean that its use produces $T_{eff}$, 
spectral energy distributions and line profiles consistent with observations. 
The precise numerical value of $\alpha_{MLT}$ we derive for a given MLT flavour, 
is tied to the specific set of input physics adopted in the model computation. 
We employ what we believe are among the best possible 
physically motivated choices for these inputs. Inaccuracies in the current generation of inputs  
for stellar models (and model atmosphere) may of course impact the calibration of $\alpha_{MLT}$.

Section ~2 summarizes the MLT formalisms and the choices of $a, b$ and $c$ in its ML2 and BV58 
flavours. A comparison of solar stellar models, low mass stars 
in the metal poor regime and intermediate mass models computed with these two prescriptions for convection 
is discussed in Sect.~3, together with the comparison with $T_{eff}$ estimates of 
RGB stars in globular clusters. Conclusions follow in Sect.~4.

\section{MLT formalisms}

As already mentioned in the Introduction, 
there are three free parameters entering the MLT equations 
\citep[denoted here as $a, b, c$ following the formalism of][]{tas90} besides 
the mixing length. 
These 3 parameters appear (together with $\alpha_{MLT}$) in the following equations for 
${\rm v_c}$, $F_c$ and $\Gamma$ 

\begin{equation}
{\rm v}_c^2=\frac{a l^2 g Q (\nabla - \nabla^{'})}{H_p}
\label{eqml21}
\end{equation}
\begin{equation}
F_c=\frac{b \rho {\rm v}_c c_p T l (\nabla - \nabla^{'})}{H_p}
\label{eqml22}
\end{equation}
\begin{equation}
\Gamma\equiv \frac{\nabla-\nabla^{'}}{\nabla^{'}-\nabla_{ad}}=\frac{c_p \rho^2 l {\rm v}_c \kappa}{c \sigma T^3}
\label{eqml23}
\end{equation}

where $\nabla^{'}$ is the temperature gradient of a rising (or falling) element of matter within the convective 
region, $\nabla$ is the average temperature gradient of all the matter at a given level 
within the convective zone (the quantity needed to solve the stellar structure equations) 
and $Q\equiv -({\rm d ln}\rho/{\rm d ln}T)_P$. 
Table~\ref{tab:ml} summarizes the BV58 and ML2 choices for $a, b, c$. 
>From these expressions for ${\rm v_c}$, $F_c$ and $\Gamma$ it is possible to obtain a simple 
algebraic equation whose solution provides the value of $\nabla$ at a given value of $r$.  

\begin{table}
\begin{tabular}{lrrr}
\hline
MLT choice   & $a$ & $b$ & $c$ \\ 
\hline
BV58	&     $\frac{1}{8}$   &	$\frac{1}{2}$ & 24\\
ML2	&	   1          &	2             & 16\\	
\hline
\end{tabular}
\caption{Values of the free parameters (besides $\alpha_{MLT}$) in the BV58 and ML2 flavours 
of the MLT.}
\label{tab:ml}
\end{table}

\citet{cg68} provides a widely used implementation of the BV58 
flavour of the MLT. Here we show how to transform their implementation to describe the ML2. 

The equation for the effective temperature gradient in \citet{cg68} is 
\begin{equation}
\zeta^{1/3}+B\zeta^{2/3}+a_0B^2\zeta-a_0B^2=0
\label{eqmlt}
\end{equation}
where $\zeta$ is defined as
$$\zeta\equiv\frac{\nabla_r-\nabla}{\nabla_r-\nabla_{ad}}$$
Once the value of $\zeta$ is computed, the knowledge of $\nabla_r$ (radiative gradient) 
and $\nabla_{ad}$ (adiabatic gradient) provides immediately the actual gradient $\nabla$.
The quantities $B$ and $a_0$ entering Eq.~\ref{eqmlt} 
are obtained through the following relationships:

$$\Gamma=A \ (\nabla-\nabla^{'})^{1/2}$$
$$\nabla_r-\nabla=a_0 \ A \ (\nabla-\nabla^{'})^{3/2}$$
$$B\equiv[(A^2/a_0) \ (\nabla_r-\nabla_{ad})]^{1/3}$$

To employ Eq.~\ref{eqmlt} with the ML2 choices of the constants $a$, $b$ and $c$ 
(Eqs.~\ref{eqml21}, \ref{eqml22}, \ref{eqml23}),  
the quantities $a_0$, $A$, and $B$ calculated with the BV58 have to be transformed as follows:

$$a_0{\rm (ML2)} =a_0{\rm (BV58)} \ \frac{8}{3}$$

$$A{\rm (ML2)}=A{\rm (BV58)} \ 3 \ \sqrt{2}$$

$$B{\rm (ML2)}=B{\rm (BV58)} \ \left(\frac{27}{4}\right)^{1/3}$$

\section{Model comparison}
\label{model}

The canonical way of calibrating $\alpha_{MLT}$ (together with the initial solar H, He and metal 
mass fractions $X$, $Y$ and $Z$) is to compute a theoretical solar model 
that reproduces -- at the solar age -- the observed solar radius, luminosity and 
$(Z/X)_{\odot}$ ratio \citep[see, e.g.][]{gou76}.
We have calibrated a theoretical solar model (including atomic diffusion) 
as described in \citet{pcsc04}, starting from the pre-Main Sequence stage. 
Our adopted stellar evolution code and state-of-the-art input physics 
are fully described in \citet[][and references therein]{pcsc04} with the exception of the 
low temperature opacities ($T \le 10^4$~K), for which we used here the updated  
calculations by \citet{fer05} that supersede the older \citet{af94} tables. 
The electron conduction opacities have also been updated, by employing the new calculations 
by \citet{cpp07}, although they are not important for the solar calibration.  
Other relevant physics inputs are 
the equation of state by A.~Irwin\footnote{http://freeeos.sourceforge.net/} 
that reproduces very well the results of the accurate OPAL \citep{rn02} 
equation of state, as discussed in \citet{pcsc04}; the OPAL radiative opacities \citep{ir96} for 
$T > 10^4$~K, and the NACRE nuclear reaction rates \citep{nacre99}.
Atomic diffusion (without the effect of radiative levitation) is treated according to \citet{tbl94} and 
the surface boundary conditions were obtained by integrating the solar $T(\tau)$ 
relationship by \citet{ks66}. We employed, again as in \citet{pcsc04}, 
the \citet{gn93} solar metal mixture, given the strong conflict \citep{bbps05} between 
inferences from helioseismology and solar models computed with the recent \citet{ags05} results. 

The solar calibration with the BV58 and ML2 provides initial abundances 
$Y$=0.2755 and $Z$=0.0201 in both cases. The actual radius of the bottom 
of the convective envelope and  
He mass fraction in the envelope were the same in both calibrations, namely 
$R_{cz}=0.7158R_\odot$ and $Y$=0.243. 
The predicted value of $Z/X$ for the present Sun is $(Z/X)$=0.0244 in both calibrations, 
in agreement with $(Z/X)_{\odot}=0.0245\pm0.005$ by \citet{gn93}.
The calibrated values of $\alpha_{MLT}$ are, respectively,    
$\alpha_{\odot, BV58}$=2.01 and $\alpha_{\odot, ML2}$=0.63. 
Notice the consistency of $\alpha_{\odot, ML2}$ with the best choice for WD atmosphere modelling. 

\realfigure{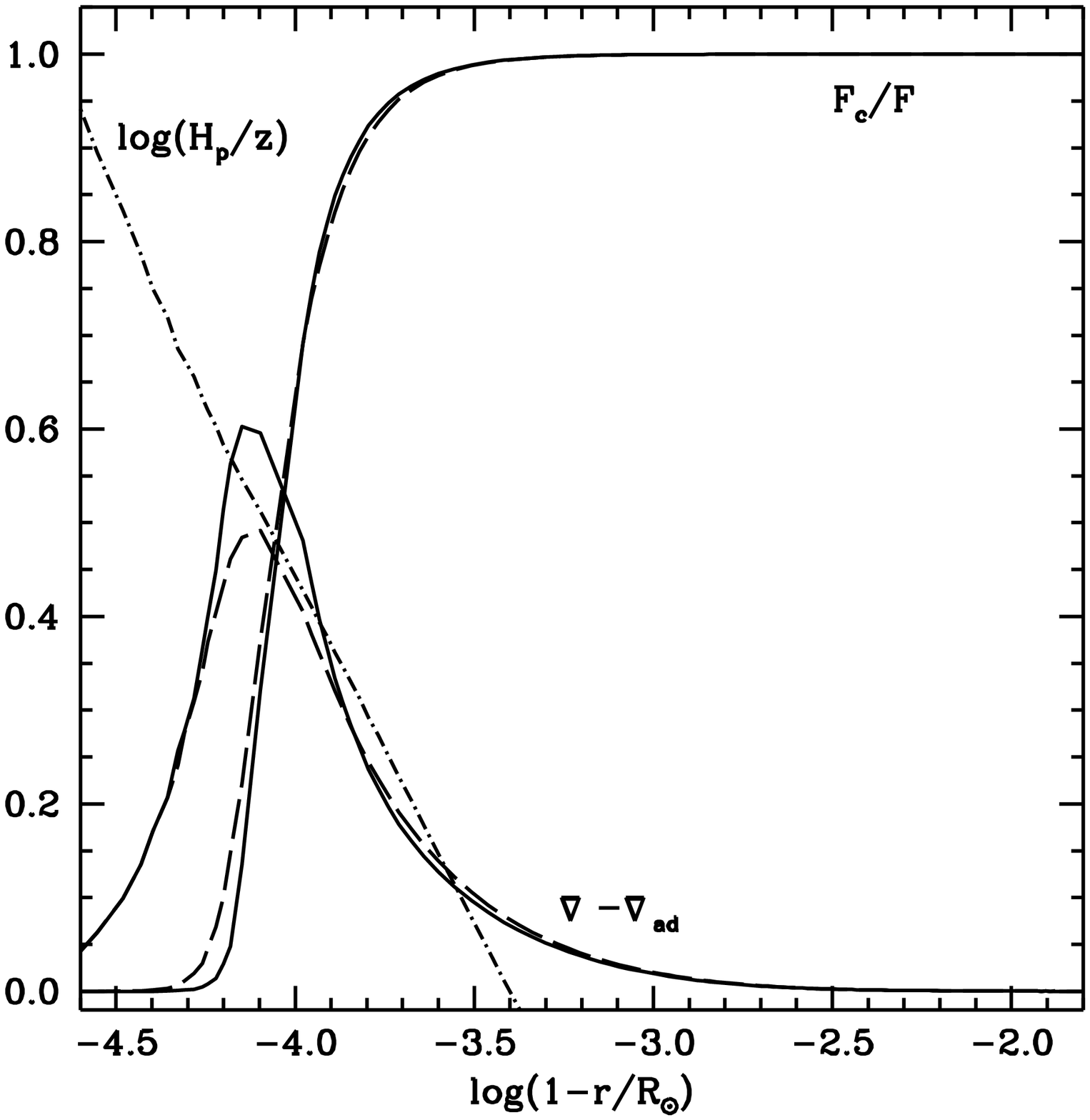}{Run of the superadiabaticity 
($\nabla$-$\nabla_{ad}$) and the ratio of the convective to the total 
energy flux as a function 
of the radial location in the outer layers of the solar convection zone. 
Solid lines represent the ML2 model, dashed lines the BV58 model.
The dashed dotted line displays the ratio between the local pressure scale height and the 
geometrical distance from the top of the convective region.   
}{f:super}

To study the properties of superadiabatic regions 
calculated with the ML2, we compare  
in the following the run of some physical variables along the solar convective envelope, with both 
the BV58 and ML2 models. 
The difference $(\nabla-\nabla_{ad})$ in the outer layers is displayed 
in Figure~\ref{f:super}, and shows how the peak of the 
superadiabaticity is higher and slightly narrower with the ML2 formulation. 
This stems from the fact that at the location of the $(\nabla-\nabla_{ad})$ peak, the 
fraction of the total energy flux carried by convection is smaller in the ML2 model (see Fig.~\ref{f:super}) 
hence a larger superadiabatic gradient is needed to satisfy the energy transport equation.
The ratio between $F_c$ and the total energy flux $F$ is also displayed and, as expected, 
differences between the two models are generally small and obviously restricted to the superadiabatic region. 
An interesting quantity also shown in 
Figure~\ref{f:super} is the ratio between the local pressure scale height and the geometrical 
distance from  the top of the convective envelope. Around the peak of the superadiabatic region the 
local value of $H_p$ is about 3-3.5 times the distance from the surface. This highlights  
a well known inconsistency \citep[e.g.][]{maz99} when using the BV58 calibrated values of $\alpha_{MLT}$, given 
that with $\alpha\sim$ 2, the mixing length $l$ in the superadiabatic region is about 6-7  
times longer than the distance $z$ from the surface. Use of the ML2 alleviates -- but does not 
eliminate -- this conflict, producing a local $l$ more comparable with $z$.  

\realfigure{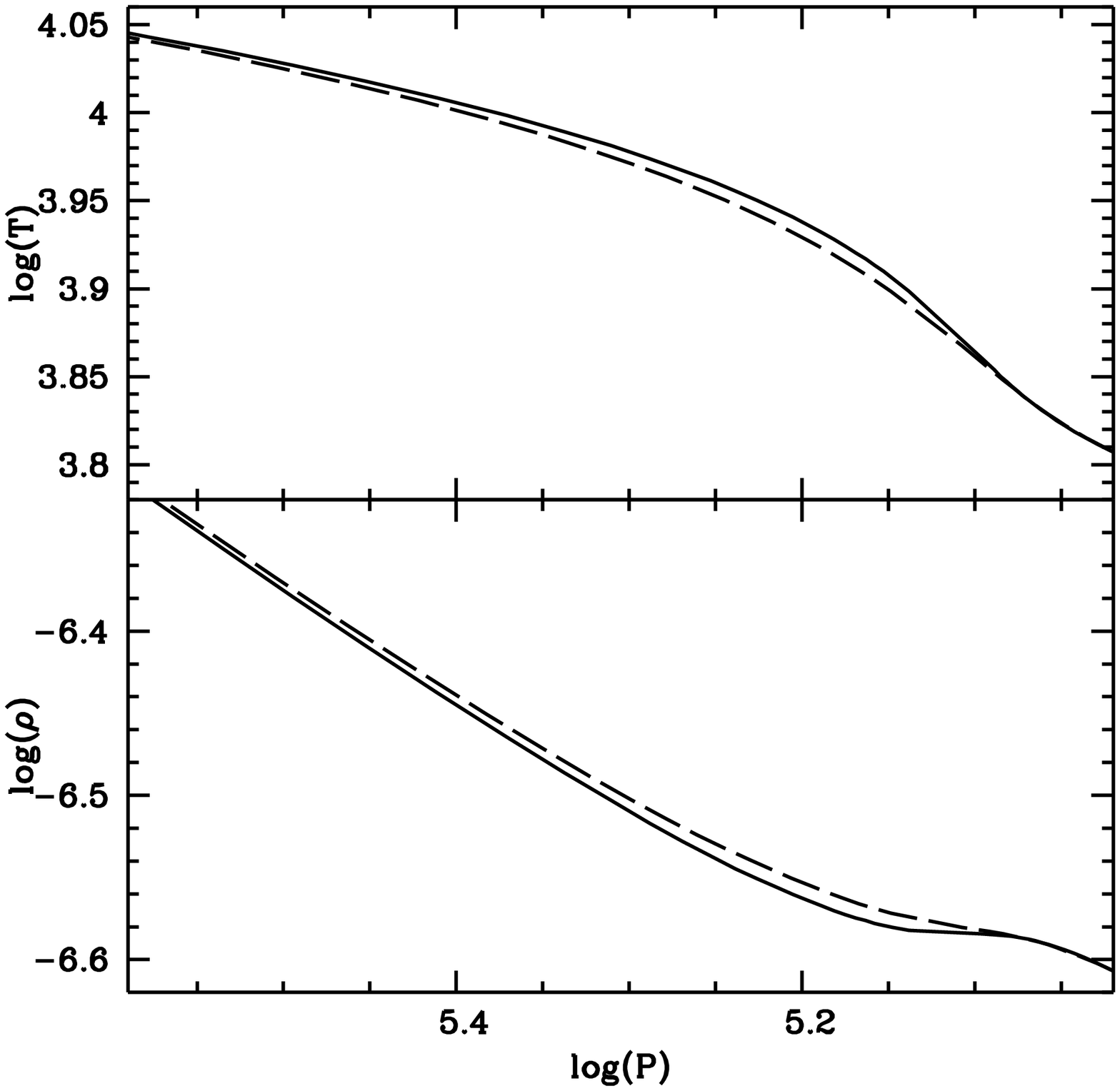}{Run of $T$ vs $P$ (upper panel) and $\rho$ vs $P$ (lower panel) 
in the outer layers of our solar models. 
The ML2 results are displayed as solid lines, the BV58 results as dashed lines.}{f:strat}

The internal $P-T$ and $P-\rho$ stratifications are displayed in Fig.~\ref{f:strat}. 
Differences in the treatment of the superadiabatic layers produce slight 
modifications in the temperature and density values at a given P, that disappear when 
moving either towards the surface or towards the inner adiabatic layers. 
The largest differences in the stratification appear 
between log($P$)$\sim$5.10 and log($P$)$\sim$5.15, close to the region of the 
superadiabatic peak. Here, the temperature in the ML2 model increases more steeply 
towards the interior than in the BV58 case, and the $\rho$ profile is almost flat.
Finally, Fig.~\ref{f:sound} shows the relative sound speed 
(${\rm v_s}=\sqrt{\frac{P \Gamma_1}{\rho}}$) difference  
along the entire structure of the two models. 
The values of ${\rm v_s}$ are essentially identical  
except in the superadiabatic region (Fig.~\ref{f:super}) as expected. 
However, the maximum difference is only $\sim$ 0.4 \% (the ML2 sound speed is
larger) just below the location of the peak of the superadiabatic gradient.
Moving inwards the ML2 sound speed becomes smaller by 0.2 \% and then equals the BV58 profile.

\realfigure{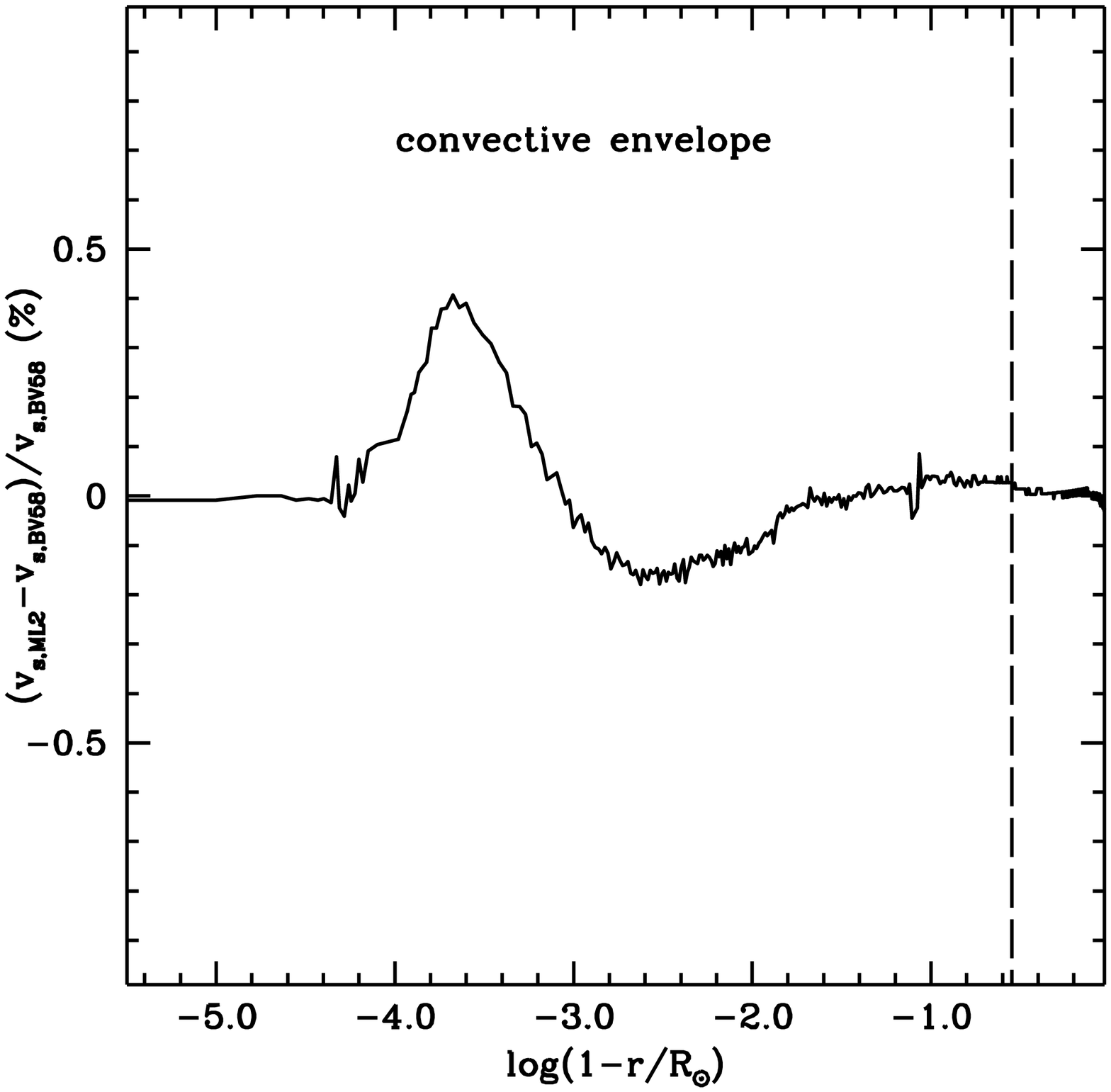}{Relative sound speed difference between the ML2 and BV58 
models, as a function of the distance from the centre. The bottom of the convective region 
is marked by a vertical dashed line.}{f:sound}

After calibrating the solar model we have computed a number of metal poor, 
low-mass models, from the pre-Main Sequence to the tip of the RGB, to 
compare solar calibrated $\alpha_{MLT}$ models in a regime different from the solar envelope 
and test predicted RGB $T_{eff}$ values against the determinations by \citet{fro83} 
in a sample of Galactic globular clusters. Globular cluster RGB $T_{eff}$ estimates are 
one of the key data to test the treatment of superadiabatic convection in 
metal-poor low-mass stars \citep[see, e.g.][and references therein]{scw02, fvso06}.

To this purpose, stellar masses with a lifetime of 12--12.5~Gyr 
have been chosen as representative of objects 
populating the Turn Off and RGB  of the observed Colour Magnitude Diagrams of Galactic 
globular clusters. We considered the $\alpha$-enhanced ([$\alpha$/Fe]=0.4) 
metal mixture and $Y-Z$ relationship in 
the \citet{pcsc06} $\alpha$-enhanced models, and computed models for [Fe/H]=$-$2.14, $-$1.62, 
$-$1.01, $-$0.60, corresponding  to $Z$=0.0003, 0.001, 0.004 and 0.01, respectively.
Sources for the input physics are the same as for the solar calibration. 
The boundary conditions for these models are still obtained by integrating the 
\citet{ks66} solar $T(\tau)$. The analysis by \citet{vdb08} has shown that boundary conditions from 
their differentially corrected model atmospheres \citep[see][for details]{vdb08}   
that employ the same MLT constants as in the stellar interiors, provide almost the same 
solar $\alpha_{MLT}$ and also very similar $T_{eff}$ along the RGB at low metallicities as 
when the \citet{ks66} $T(\tau)$ relationship is used.

Figure~\ref{f:RGB1} displays the tracks for [Fe/H]=$-$2.14 ($M=0.8~M_{\odot}$) 
and [Fe/H]=$-$0.60 ($M=0.92~M_{\odot}$) from the pre-Main Sequence to the tip of the RGB. 
The ML2 and BV58 tracks at the same [Fe/H] are almost identical. 
Differences in $T_{eff}$ between the two sets of tracks begin to appear at the base 
of the RGB and slowly increase towards the RGB tip, when the superadiabatic region 
is located at lower temperatures and lower densities compared to previous evolutionary phases. 
In quantitative terms, they amount at most to $\sim$40~K 
for [Fe/H]=$-$2.14, and $\sim$50~K for [Fe/H]=$-$0.60. 
The luminosity of the RGB bump is the same, irrespectively 
of the MLT flavour, and also the amount of He dredged up along the RGB is the same. This is a  
consequence of the agreement between the location (in mass) of the bottom boundary of 
the surface convective regions in both sets of models, all the way to the tip of the RGB. 
Evolutionary timescales are unaffected by the change of the MLT flavour.

\realfigure{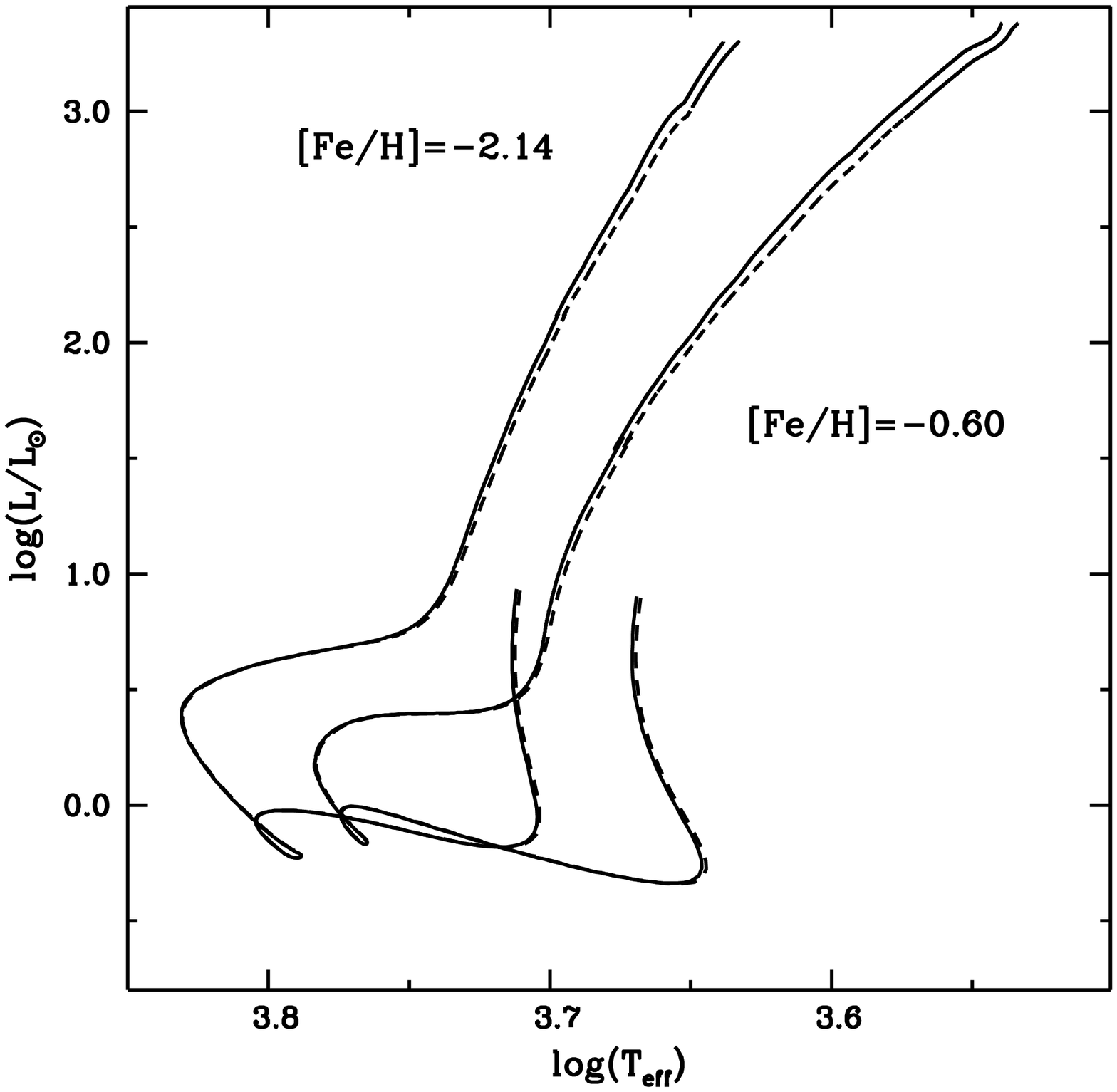}{Evolutionary tracks of low-mass models for the labelled [Fe/H] values and an 
$\alpha$-enhanced metal distribution, 
from the pre-Main Sequence to the tip of the RGB, computed with the 
solar calibrated BV58 (dashed lines) and ML2 (solid lines) flavours of the MLT (see text for details)}{f:RGB1}

\realfigure{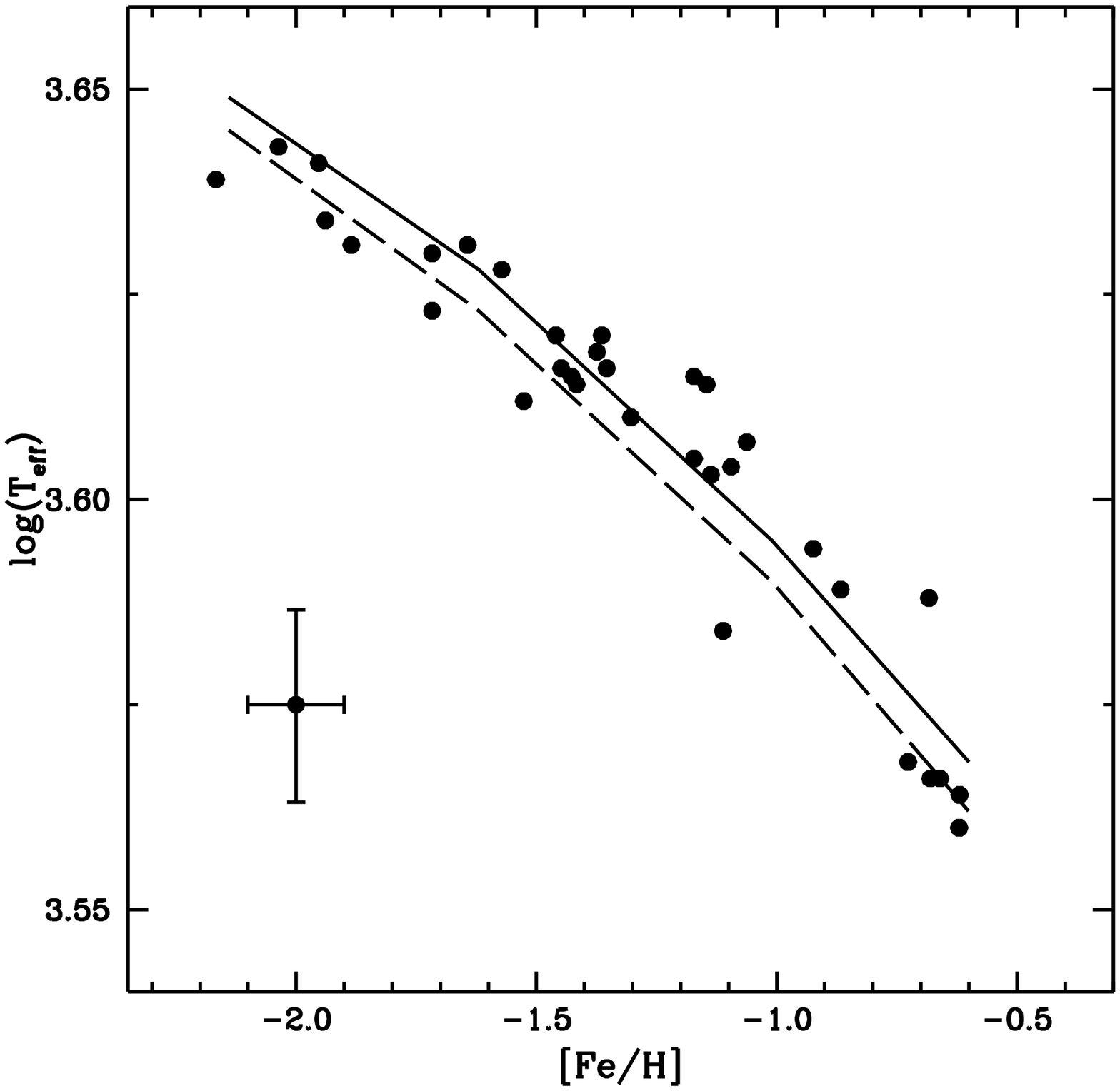}{Estimates of the average RGB $T_{eff}$ at $M_{bol}=-$3 
for a sample of Galactic globular clusters, compared to the corresponding values 
from our models, computed with the solar calibrated BV58 (dashed line) and ML2 
(solid line) flavours of the MLT. The assumed typical error bars 
are displayed in the lower left corner (see text for details).}{f:RGB2}

Figure~\ref{f:RGB2} compares also estimates of 
the mean $T_{eff}$ at $M_{bol}$=$-$3.0 for the RGBs of a sample of Galactic globular 
clusters by \citet{fro83}, with the corresponding quantities from our models. Theoretical 
bolometric luminosities have been transformed into bolometric magnitudes assuming 
$M_{bol, \odot}$=4.75, to have the same zero point of \citet{fro83} bolometric magnitudes. 
The cluster [Fe/H] values are on the \citet{cg97} scale. We assigned an 0.1~dex error bar 
to the cluster [Fe/H] values, and a $\pm$100~K error to the temperatures \citep[as an estimate 
of realistic systematic errors in the RGB temperature scale, see e.g.][]{aam99}.

Both MLT flavours produce $T_{eff}$ values that 
appear consistent with \citet{fro83} data. 
We computed the differences between the predicted and estimated $T_{eff}$ for each cluster, 
by interpolating (with a cubic spline) 
among our results for the individual cluster metallicites. We did not find any 
statistical significant correlation 
between the $T_{eff}$ differences (observations-theory) and 
the corresponding [Fe/H], for both ML2 and BV58 
models. In case of the ML2 models the average difference 
(observations-theory) is equal to $-$15 ~K, with a rms of 65~K. The 
same difference for the BV58 models is equal to +32~K, with a rms of 65~K. The agreement 
with observations appear good in both cases. The smallest difference is for the ML2 results, 
but both sets of models provide $T_{eff}$ consistent with \citet{fro83} if one considers 
the rms dispersion around this mean values  
and the estimated systematic errors on the RGB $T_{eff}$ scale.

We remark here that these models, at 
variance with the solar calibration, have been computed without including the effect 
of atomic diffusion, whose efficiency in Galactic globular cluster stars is still 
debated \citep{gbb01, kor06}. In any case, 
the efficiency of diffusion does not affect differential comparisons 
along the Main Sequence, 
because of the same mass extension of convective envelopes in both ML2 and BV58 models (hence 
the same rate of depletion of surface He and metals in both cases, and the same 
effect on $T_{eff}$). Also,  
the comparison with estimated $T_{eff}$ values of RGB stars is robust, given that the 
location of the RGB in the H-R diagram 
is largely insensitive to the efficiency of diffusion.

\realfigure{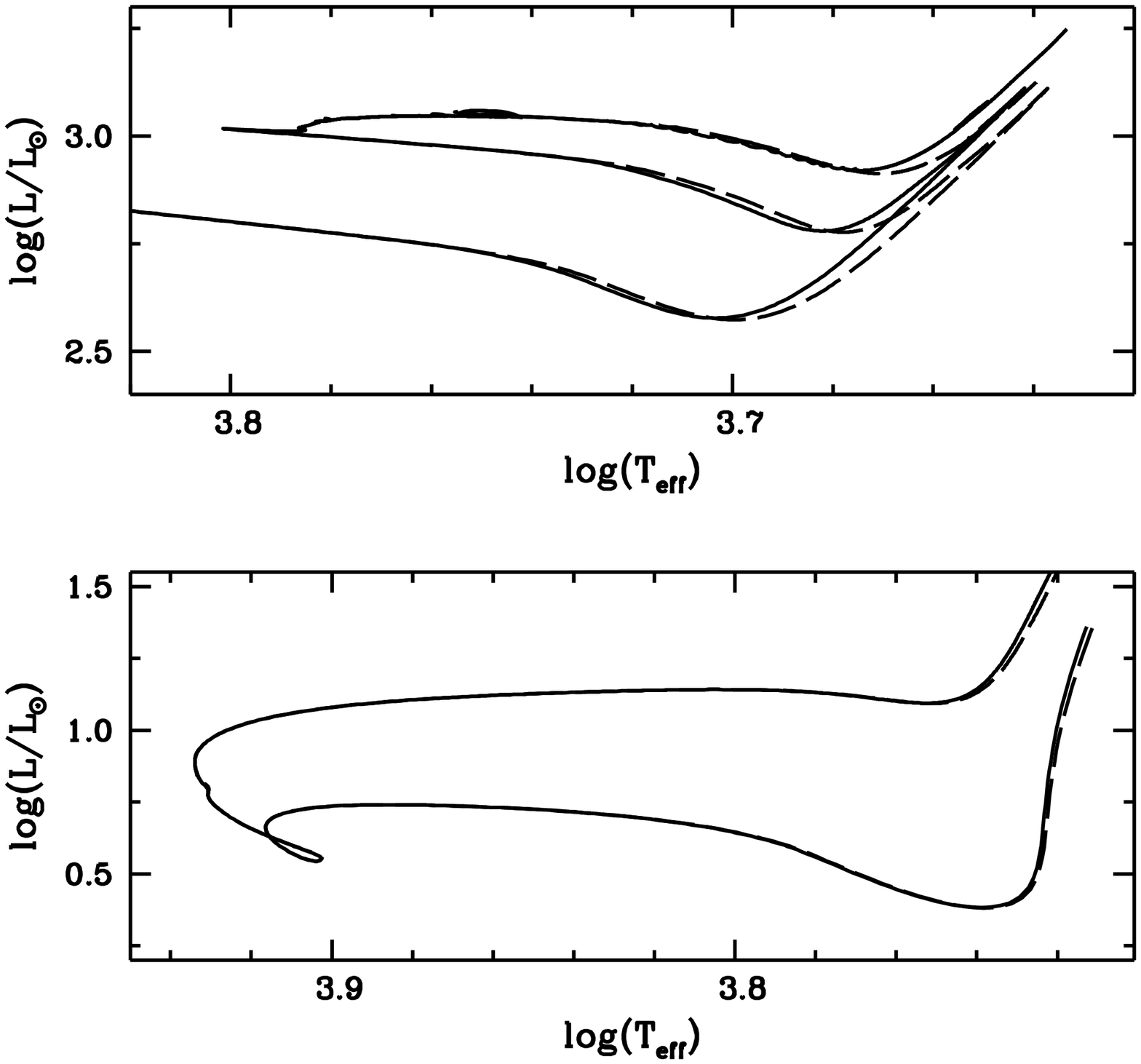}{Evolutionary tracks for, respectively, a 5$M_{\odot}$ star with 
solar chemical composition 
(from the subgiant branch to the early Asymptotic Giant Branch phase -- upper panel) 
and a 1.2$M_{\odot}$ star with [$\alpha$/Fe]=0.4 and [Fe/H]=$-$1.62 (from the pre-Main Sequence to the base 
of the RGB -- lower panel). As in the previous figures, solid lines display the 
ML2 models, dashed lines the BV58 models.}{f:small}

As a final test, we display in Fig.~\ref{f:small} models for a 1.2$M_{\odot}$ star with 
an $\alpha$-enhanced, $Z$=0.001 ([Fe/H]=$-$1.62) metal composition, 
and a 5$M_{\odot}$ star with solar composition, with both solar calibrated BV58 and ML2 convection. 
Both objects display during their evolution shallow or vanishing surface convective regions, and allow us to 
compare the two MLT flavours in the regime of very thin convective envelopes. 
The ML2 and BV58 calibrations produce again identical tracks along the Main Sequence and 
turn off region of the 1.2$M_{\odot}$ star, where convection is efficient in a very shallow outer layer. 
Differences in $T_{eff}$ along the pre-Main Sequence and RGB 
are as for the lower mass objects discussed above. 
Evolutionary tracks with the BV58 and ML2 for the 5$M_{\odot}$ solar composition model are also shown in 
Fig.~\ref{f:small}. We have displayed the post-Main Sequence 
evolutionary phases where surface convection appears in the models. 
Differences in $T_{eff}$ are again small, reaching at most 50~K along the bright RGB 
and the early Asymptotic Giant Branch.
The amount of He dredged up during the post-Main Sequence phases is the same with 
both treatments of surface convection, as a consequence of equal depths of the convective regions.

\section{Conclusions}

In the previous section we have shown that 
solar calibrated BV58 and ML2 stellar models  
computed with the input physics and boundary conditions described in Sect.~\ref{model}, 
are very similar. There are some differences in the physical structures of the superadiabatic 
layers -- as expected -- together with minor variations in $T_{eff}$, 
but the extension of the convective regions are the same. 
Effective temperature differences appear essentially at the RGB and slowly increase 
towards the RGB tip, reaching at 
most 40-50~K at the lowest temperatures. This latter result does not conform exactly to 
\citet{ped90} conclusion that models computed with different choices for the 
MLT parameters $a$, $b$ and $c$ should be equivalent once $\alpha_{MLT}$ is recalibrated, 
but the differences in the RGB $T_{eff}$ are not large. 

Both ML2 and BV58 solar calibrated models provide a good match to $T_{eff}$ 
estimates for bright RGB stars in Galactic globular clusters. ML2 models appear to perform 
marginally better when observation-theory differences for each cluster are compared, 
but the effect of changing the MLT flavour is very small compared 
to both the rms scatter of the individual values, and realistic estimates of systematic 
errors in the RGB $T_{eff}$ scale.
 
In solar models, we have compared in some detail the  
sound speed, T, $\rho$ stratifications in the convective layers. If the ML2 
improves the agreement between predicted and observed helioseismological 
properties \citep[e.g. through the comparison of the solar $p$ mode excitation rates, see][]{sk06} 
is a matter to be established. On the other hand, due to its crudeness, we do not expect the MLT to be 
the adequate tool to model local details of superadiabatic convection 
to the degree of accuracy needed to satisfy helioseismological constraints on the 
structure of the solar superadiabatic layers.

A crucial point is that $\alpha_{MLT}$ in the ML2 solar calibration obtained with our 
code and the adopted state-of-the-art stellar input physics is equal to 0.63, a 
value consistent with the calibration preferred in WD model atmosphere calculations, that deal with 
a very different regime of gas properties. 
The exact value 
of the solar $\alpha_{MLT}$ may surely be subject to some variations, when different sources for the input 
physics are employed, but we expect that with nowadays best (i.e. most accurate and updated) possible choices, 
values around $\alpha_{MLT}\sim 0.6$ should still be found. 

To verify whether this unification of the MLT treatment can be pushed even further, it will be 
necessary -- as a next step --  to compute and test model atmospheres (plus spectra and bolometric corrections) for the 
pre-WD phases with surface convection, using the ML2 formalism with $\alpha_{MLT}\sim$0.6. 
These ML2-based model atmospheres, when computed and as a further test of this unified MLT treatment, 
should then be also employed to determine the boundary conditions 
of interior models, for studying the consistency with results obtained with a (solar) $T(\tau)$ integration.


\begin{acknowledgements}
S.C. acknowledges the financial support of INAF through the 
PRIN 2007 grant n. CRA 1.06.10.04 \lq{The local route to galaxy
formation: tracing the relics of the hierarchical merging process in the Milky Way and in other nearby galaxies}\rq. 
We warmly thank S. Degl'Innocenti for interesting discussions on the Solar Standard Model and Helioseismological constraints, 
and our referee (A. Dotter) for a prompt report and comments/suggestions that helped to 
improve the presentation of our results.

\end{acknowledgements}

\bibliography{bib_0253}
\bibliographystyle{aa}

\end{document}